\documentclass[12pt]{article}
\usepackage{amsmath}
\usepackage{graphicx,psfrag,epsf}
\usepackage{enumerate}
\usepackage{natbib}
\usepackage{bm}
\usepackage{multirow}

\begin{document}

\def\spacingset#1{\renewcommand{\baselinestretch}%
{#1}\small\normalsize} \spacingset{1}


  \title{\bf A Comprehensive Method for Solving Finite-State Semi-Markov Processes}
  
  \author{Richard L. Warr \footnote{The views expressed in this article are those of the author and do not reflect the official policy or position of the United States Air Force, Department of Defense, or the U.S. Government.} \\
    Department of Mathematics and Statistics \\ Air Force Institute of Technology \\ Wright-Patterson Air Force Base, OH 45433 \\ (richard.warr@afit.edu)\\
    and \\
    David H. Collins \footnote{This work performed under the auspices of the Los Alamos National Security, LLC, for the National Nuclear Security Administration of the U.S. Department of Energy under contract DE-AC52-06NA25396. } \\
    Statistical Sciences Group \\ Los Alamos National Laboratory \\ Los Alamos, NM 87545 \\ (dcollins@lanl.gov)}
  \maketitle

\bigskip
\begin{abstract}
Semi-Markov processes (SMPs) provide a rich framework for many real-world problems.  However, due to difficulty implementing practical solutions they are rarely used with their full capability.  The theory of SMPs is quite mature but was mainly developed at a time when computational resources were not widely available.  With the exception of some of the simplest cases, solutions to SMPs are inherently numerical, and SMPs have been underutilized by practitioners because of difficulty implementing the theory in applications.  
This paper demonstrates the theory and computational methods needed to implement SMP models in practical settings.  Methods are illustrated with an application modeling the movement of coronary patients in a hospital.  Our aim is to allow practitioners to use richer SMP models without being burdened with the rigorous mathematical theory. 

\end{abstract}

\noindent%

\newpage


\section{Introduction}
Semi-Markov processes (SMPs) provide a rich framework for many real-world problems.  SMPs include Markov processes, Markov chains, renewal processes, Markov renewal processes, Poisson processes, birth and death processes, and M/G/1 queues to name a few.  They have been applied in the areas of survival analysis \cite[]{StocNetModels}, reliability \cite[]{SemiMarkov}, DNA analysis \cite[]{barbu}, queuing theory \cite[]{Kleinrock:1}, finance \cite[]{Janssen:2}, and many others.  However, SMPs have not been used as widely as one would expect, given their generality.  A key reason for this is the perceived complexity of solving semi-Markov models.  Many practitioners prefer ``simpler" models such as continuous-time Markov chains, but (as we demonstrate in this paper) solving semi-Markov models for quantities of interest is in fact straightforward.  Within the realm of finite-state processes, SMPs are extremely flexible, and we believe they should be a primary statistical tool for modeling stochastic processes.

The concept of semi-Markov processes is generally agreed to have been independently introduced by \cite{levy54}, \cite{Takacs54}, and \cite{WLSmith}.  The theory was formalized in \cite{pyke1} and \cite{pyke2}.  Further developments came from \cite{Takacs59}, \cite{PykeScha1}, \cite{PykeScha2}, and \cite{Cinlar}.  Since that time there have been many contributions to and applications of SMPs; for typical applications, however, we feel that the key problems in SMPs were solved many years ago, and the relevant theory is treated comprehensively in the cited papers by Pyke.  Pyke's work provided general solutions in terms of Laplace transforms, however, the actual quantities of interest could not be obtained without numerical computation to recover probability distributions. The necessary numerical algorithms and computational power needed to solve SMPs are readily available today (as we will show), but were not within reach for most researchers in the early 1960s when the major results for SMPs were published.  Therefore, the bridge from theoretical solutions to practical implementation was not immediately established.  Statistical flowgraph models (SFGMs), as detailed in \cite{flowgraphs}, were a beginning in this endeavor.  SFGMs have primarily focused on finding the first passage distribution from a beginning state to some absorbing final state, using methods developed in \cite{mason53}.  This paper more fully explores connecting the theory of SMPs with computational methods.  In particular, solutions for several key quantities found in \cite{pyke2} are implemented using the R software program without the need for any additional packages or add-ons.  However, these calculations could be performed in any program with numeric integration and matrix manipulation capabilities, e.g. MATLAB, Mathematica, and FORTRAN.

So far we have used the term ``solve" ambiguously; more precisely, it is the process whereby we calculate certain quantities from an SMP.  Some of the quantities desired from a time-homogeneous finite-state SMP, given the process was started in state $i$ at time $t=0$, are:

\begin{enumerate}
\item the probability the process is in state $j$ (as a function of time),
\item the first passage distribution of the time to reach state $j$,
\item the probability of reaching a state $j$, $k$ number of times (as a function of time),
\item the probability of reaching a state $j$, $k$ or fewer times (as a function of time),
\item the expected number of times the process has been in state $j$ (as a function of time),
\item the long run probability that the process will be in state $j$.
\end{enumerate}
\label{SMP-itemList}

These are significant quantities that modelers find difficult to compute for even simpler Markov processes, but the theory and computational techniques currently exist to solve them in general terms for SMPs.  With today's computational resources, processes that have many states with any smooth transition distributions can be solved in a matter of seconds.  In the remainder of the paper, we detail how to find these quantities.

In our experience, the primary focus in stochastic processes has been finding asymptotic probabilities.  However, there is a considerable need for finding time-dependent state probabilities of a process, before it is asymptotically ``well behaved".  For example, if we were developing a business model it would be very reasonable for investors to want to know the probability of bankruptcy at 1 month, or 2 years into the future.  These probabilities, as time goes to infinity, are of little consequence.

\begin{figure}[ht]
\begin{center}
\includegraphics[width=4in]{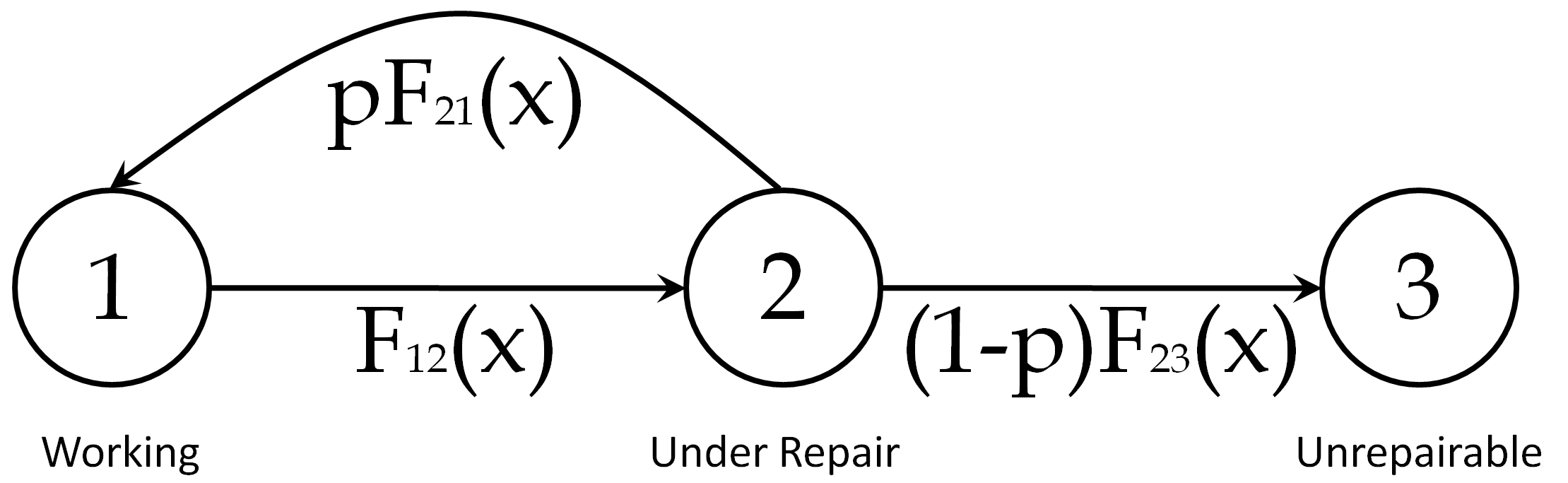}
\caption{A reliability semi-Markov process.}
\label{figRelExamp}
\end{center}
\end{figure}

As another example, the reliability model in Figure \ref{figRelExamp} has a single absorbing state (state 3), so the asymptotic probabilities are trivial: 1 for the absorbing state, and 0 for the other (transient) states. In the context of this reliability example it would be very valuable to find the six numbered quantities listed above, i.e., to know how the process behaves before reaching state 3.   If the system happened to be an automobile, then it would be worth knowing the probability that a vehicle is in the ``Working" or ``Under Repair" states, as well as calculating the first passage distribution to the ``Unrepairable" state. Other items such as the probability of $k$ repairs before time $t$ or the expected number of repairs up to time $t$ could be used for developing efficient manufacturer warranties.

\section{Background}

A real deterrent to using SMPs is the effort required to decipher the notation. In this section we define the notation used in this paper as a reference for the reader, and list assumed properties of the models we deal with.  We have attempted to conform to the SMP literature as closely as possible.  We also introduce Laplace transforms, give some of the properties that make them useful, and show how to numerically invert them into probability density functions (PDFs) or cumulative distribution functions (CDFs).  We avoid going into measure theoretic details to make the paper as accessible as possible, so we assume absolutely continuous distributions.  However, the SMP results here have been proven using general probability measures; see \cite{pyke1}.  

\subsection{Notation and assumptions}

We assume a finite-state SMP with $n$ states running in continuous time. Roughly, this means that transitions between states constitute a Markov chain with transition probability matrix  $[p_{ij}]$; however, waiting times (times between transitions) may have arbitrary distributions depending on both the origin and destination states for the transition. This may be clearer if we contrast an SMP with a discrete-state Markov process: once the current state of a Markov process is known, no additional information about its future is provided by its past history; this implies that waiting time distributions must be memoryless (geometric in discrete time, or exponential in continuous time), and independent of the destination state. In a semi-Markov process, since waiting time distributions may be arbitrary, knowledge of how long the process has waited in its current state can provide information on how much longer it will wait, and what the destination state will be.

We also make the following assumptions: the process is time-homogeneous, i.e., the waiting time distribution for a transition $i \rightarrow j$ does not explicitly depend on how long the process has been running; there are no instantaneous transitions from a state, which implies that the number of transitions in any finite time period is finite; and there are no transitions from a state to itself. The last assumption can be made without loss of generality by transforming to an equivalent SMP without self-transitions \cite[Lemma 3.2]{pyke1}.

Important classes of states in SMPs are \textit{absorbing} states, which, once entered, are never left;  \textit{transient} states, which with probability one are eventually left and never returned to; and \textit{recurrent} states, which with probability one are eventually entered, starting at any state in the process. Many applications in survival and reliability analysis have one or more absorbing states, representing failure or death, with all other states transient. Other application areas, such as queuing and inventory processes, typically have all states recurrent.

The random variable $X$ represents occupancy time in a state, as distinguished from $T$, the time since the process was started or ``calendar time".  Most frequently we are interested in finding functions of the calendar time $t$, and unless explicitly stated, when we use the word ``time", we are referring to calendar time.  Next, let $Z(t)$ denote the state of the SMP at time $t$.  $Z(t)$ can take on any positive integer value up to and including the number of defined states in the process.

Under the assumptions given above, there is a one-to-one correspondence between the SMP $Z(t)$ and an equivalent Markov renewal process (MRP) $\textbf{N}(t)=(N_1(t),\ldots,N_n(t))$, where each $N_{_j}(t)$ counts the number of transitions into state $j$ that have occurred up to time $t$ \cite[]{pyke2}. It turns out that it is more natural to think of many of the quantities associated with SMPs in terms of the counting processes $N_{_j}(t)$. For an introductory discussion of MRPs in relation to SMPs, see \cite{RossOpt}.

The table below shows the items necessary to define a SMP:

\begin{center}
\begin{tabular}{  c | l  } \hline
  $n$ & the number of states in the SMP  \\ \hline
  $F_{_{ij}}(x)$ & the CDF of the waiting time distribution from state $i$ to state $j$ \\ \hline
  $p_{_{ij}}$ & the probability that the next state in the process is $j$, given the \\
           & process entered state $i$   \\   \hline
\end{tabular}
\end{center}

Constraints on $p_{_{ij}}$ are $\sum_{j=1}^{n}p_{_{ij}}=1$ (or equal to 0 if $i$ is an absorbing state), and all $p_{_{ij}} \geq 0$.  A boldface symbol without indices refers to a matrix valued function (e.g., $\textbf{F}(t)$ is a matrix with elements $F_{_{ij}}(t)$ in the i$^{th}$ row and j$^{th}$ column).  The Laplace Transform (LT) of a function is denoted with a tilde, so, e.g., $\tilde{F}_{_{ij}}(s)$ is the LT of $F_{_{ij}}(t)$. Similarly $\tilde{\textbf{F}}(s)$ is the matrix valued function where the LTs are of the individual elements.  We will sometimes drop the $(x)$, $(t)$, or $(s)$ on functions to reduce notational clutter.  For example, one can read $\tilde{\textbf{F}}$ as $\tilde{\textbf{F}}(s)$.  It is worth mentioning that $\textbf{F}(x)$ is in general not symmetric, so $F_{_{ij}}(x) \neq F_{_{ji}}(x)$.

Now that we have detailed what is needed to define a SMP, we formally define the quantities and functions we want to obtain from a SMP.  These are listed in the same order as in the Introduction (p. \pageref{SMP-itemList}) with a few additions; they are all conditioned on a given starting state $i$:

\begin{center}
\begin{tabular}{  c | l  } \hline
  \multirow{2}{*}{$P_{_{ij}}(t)$}  & $P(Z(t)=j|Z(0)=i)$  \\
                                   & the probability the process is in state $j$ at time $t$   \\    \hline
  \multirow{2}{*}{$G_{_{ij}}(t)$}  & $P(N_{_j}(t)>0|Z(0)=i)$  \\
  								   & the distribution of the time of first passage to state $j$    \\    \hline
  \multirow{2}{*}{$v_{_{ij}}(k;t)$} & $P(N_{_j}(t)=k|Z(0)=i)$  \\
 					& the probability of reaching a state $j$, $k$ number of times by time $t$   \\    \hline
  \multirow{2}{*}{$V_{_{ij}}(k;t)$} & $P(N_{_j}(t)\leq k|Z(0)=i)$  \\
					& the probability of reaching a state $j$, $k$ or fewer times by time $t$ \\    \hline
  \multirow{2}{*}{$M_{_{ij}}(t)$}   & $E[N_{_j}(t)|Z(0)=i]$  \\
  					& the expected number of times the process has been in state $j$ at time $t$ \\    \hline
  \multirow{2}{*}{$\pi_{_{ij}}$}    & $\lim_{ t\to\infty } P(Z(t)=j|Z(0)=i)$  \\
  					& the long run probability that the process will be in state $j$  \\    \hline
  $Z(t)$    &   An integer value denoting the state of the process at time $t$  \\    \hline
  $N_{_{j}}(t)$    &  The number of times the process transitioned to state $j$ since $t=0$  \\    \hline
\end{tabular}
\end{center}

For notational convenience, when the derivatives exist we also define:
\begin{equation} f_{_{ij}}(x) = \frac{d}{dx}F_{_{ij}}(x) \end{equation}
\begin{equation} g_{_{ij}}(t) = \frac{d}{dt}G_{_{ij}}(t) \end{equation}
\begin{equation} q_{_{ij}}(x) = p_{_{ij}}f_{_{ij}}(x) \end{equation}
\begin{equation}
  \delta_{_{ij}} = \left\{
  \begin{array}{l l}
    0 & \quad \text{if $i \neq j$}\\
    1 & \quad \text{if $i=j$}\\
  \end{array} \right.
\end{equation}
\begin{equation} h_{_{ij}}(x) = \delta_{_{ij}}\sum_{j=1}^{n}q_{_{ij}}(x) \end{equation}

$\textbf{I}$ is defined to be the $n \times n$ identity matrix, and $\textbf{1}$ to be the $n \times n$ matrix where every element is 1. The matrix operation ``$\circ$" such that $\big[\textbf{A} \circ \textbf{B}\big]_{_{ij}}=A_{_{ij}}B_{_{ij}}$ is the elementwise product of two matrices (the Hadamard product).

\subsection{Laplace transform inversion}\label{sectTransformInversion}

In the statistical community there has been considerable avoidance of transforms, even though transforms (plus the computations involved in moving between the time and transform domains) are what actually simplifies the problem of solving SMPs by making it similar to solving a system of linear equations.

A prevalent opinion among researchers is exemplified by this quote from \cite{Janssen:1}: ``. . . it is then necessary to invert Laplace transforms to obtain the solution of the equation and unfortunately, it is well known that this inverse transform is not stable numerically . . ." (p. 92); they cite \cite{Bellman:1} and \cite{Csenki:1} as support. Indeed, it has been well-known at least since the book by Bellman et al. that the {\em general} problem of Laplace transform inversion is ill-posed, and the difficulty of inverting transforms of general functions has been much discussed since then \cite[]{Davies:1,Duffy:1}.

What the conventional wisdom misses is that in analyzing SMPs we are interested only in inverting transforms of probability densities and distribution functions of non-negative random variables, functions with very special properties, and it has long been known that this restriction significantly simplifies the problem, particularly in the case of absolutely continuous distributions \cite[]{Dubner:1,abate,Straw:1}. Our own experience confirms that methods based on \cite{abate,abate2} are strikingly accurate for numerical inversion of transforms specified either in closed form or numerically. These methods are also simple to implement and require minimal tuning, in contrast to algorithms such as the Weeks method, based on orthogonal expansion in Laguerre polynomials \cite[Sect. 10.2]{Csenki:1}. Moreover, LTs need not be known in closed form if we are able to compute them by numerical integration.

The Laplace Transform of a generic function $\varphi(\xi)$ is defined as:
\begin{equation}
\tilde{\varphi}(s) = \int_{0}^{\infty} e^{-s \xi}\varphi(\xi) \, d\xi, \text{ where $s$ is a complex number.}
\label{eqn66}
\end{equation}
A useful identity for numerically calculating LTs is Euler's formula (not to be confused with the EULER method in this paper): $e^{i\theta}=\cos(\theta)+i \sin(\theta)$, where $i=\sqrt{-1}$.  If we let $s=x+iy$, Equation (\ref{eqn66}) then becomes:
\begin{equation}
\tilde{\varphi}(s) = \int_{0}^{\infty} e^{-x \xi} \cos(y \xi) \varphi(\xi) \, d\xi - i \int_{0}^{\infty} e^{-x \xi} \sin(y \xi) \varphi(\xi) \, d\xi \, ,
\label{eqn67}
\end{equation}
which enables us to calculate the LT without any numeric complex integration.
The inverse Laplace Transform can be defined by the Bromwich contour integral, where the complex contour can be any vertical line $s=a$ such that $\tilde{\varphi}(s)$ has no singularities on or to the right of it, see \cite{abate2}.  This integral is:
\begin{equation}
\varphi(\xi) = \frac{1}{2 \pi i}\int_{a-i\infty}^{a+i\infty} e^{s \xi}  \tilde{\varphi}(s) \, ds 
\label{eqn67a}
\end{equation}
Again we emphasize that we are avoiding some measure theoretic details, so although for any CDF $F$, $f_{_{ij}}(x)$ may not exist, the LT $\tilde{f}_{_{ij}}(s) = \int_0^{\infty}e^{-sx}dF(x)$ is always defined and uniquely represents that random variable \cite[Chapter XIII]{Feller:2}.

A fundamental property of the LT, and the basis of its usefulness for solving stochastic processes, is the fact that it maps convolution to multiplication. Thus if the random variables $X$ and $Y$ have densities $f_X$ and $f_Y$, the LT of the density of their sum is $\tilde{f_X}\tilde{f_Y}$. This allows complex integral equations to be solved
algebraically in the transform domain. Since the density $f(x)=F'(x)$, another well-known property of LTs that is used extensively in this paper is
\begin{equation}
\tilde{F}(s) = \frac{1}{s} \tilde{f}(s).
\end{equation}

LTs have very useful properties, but these come at a price.  Many parametric distributions do not have closed-form LTs, and even if they did, after manipulation in the LT domain they most likely would not have closed-form inverses.  With today's computational power and algorithms, however, this is no longer the significant obstacle it once was.

There are various ways to numerically invert LTs.  The methods we apply in this paper work well in terms of speed and accuracy for most distributions; this is an active area of research, and we expect to see improvements in the future. In this paper we use the EULER method as shown in \cite{abate2}.  EULER is specially suited for inverting LTs of probability distributions that are smooth (i.e., the CDF has at least two continuous derivatives).  Therefore to accurately invert LTs of non-smooth distributions other techniques should be considered.  Other possibilities include using a modification to the EULER method such as in \cite{ModEULER}, or using a discrete approximation with the use of the inverse discrete Fourier transform.

In simple terms, the EULER method approximates the exact LT inversion integral \cite[]{Doet:1} by the trapezoidal method; the result is a Fourier cosine series approximation, and by a careful choice of the step size the cosine terms become $\pm 1$. The series is then summed using Euler summation, which gives the method its name. For a function $\varphi(t)$ and its LT $\tilde{\varphi}(t)$, the result is

\begin{equation}
\varphi(t) \approx \sum_{j=0}^N \, (-1)^j \, w_j \, \text{Re}\Bigg[\tilde{\varphi}\Big(\frac{A}{2t}+ \frac{j\pi i}{t}\Big)\Bigg],
\label{eqn9}
\end{equation}
where $Re[\cdot]$ gives the real portion of a function and $w_j$ is a weight associated with each term (for Euler summation the weights are binomial coefficients). $N$, $A$, and the $w_j$'s control the accuracy of the approximation; see \cite{abate} for details.

%

\section{Semi-Markov Process Theory}\label{sectTheory}

Now that we have introduced the terminology and methods of computation we move to the main focus of the paper.  Figure \ref{figRelExamp} shows a simple example of a semi-Markov process that might be encountered in a reliability application. States 1 and 2 are transient, and state 3 is absorbing. Here we might be interested in quantities such as the number of times the process is in state 1 in a 5 year period, the distribution of the time until the process reaches state 3, or the average time spent in state 1 from 0 to 3 years.  In this section we show how these questions and others can be answered, by finding the LTs of the quantities $P_{_{ij}}(t)$, $G_{_{ij}}(t)$, $v_{_{ij}}(k;t)$, $V_{_{ij}}(k;t)$, and $M_{_{ij}}(t)$.

\subsection{Time-dependent state probabilities}

The time-dependent state probabilities $P_{_{ij}}(t)$ are invaluable pieces of information from a multi-state model.  For example, in a reliability application, time-dependent state probabilities could provide probabilities of being in an ``Unrepairable" or ``Working" state at a given moment in time.  At the beginning of a process these probabilities might be significantly different from the long run probabilities.  The formula to find these probabilities is found in Equation (4.1) in \cite{pyke2}, which is
\begin{equation}
\tilde{\textbf{P}}(s) = \frac{1}{s}\Big(\textbf{I}- \tilde{\textbf{q}}(s) \Big)^{-1} \Big(\textbf{I}- \tilde{\textbf{h}}(s) \Big).
\end{equation}
Given the LTs of these probabilities, the probabilities themselves can be found by inversion back to the time domain.  The actual computation will be demonstrated in the application section.

Another quantity of interest that is related to $P_{_{ij}}(t)$ is the expected amount of time the process will be in a state during a specific period of time.  For example in the reliability example in Figure \ref{figRelExamp}, it would be of interest to determine the expected amount of time the system is in ``In repair" during some time interval.
The expectation of time in state $j$ in the interval $[0,t_1)$ is $\int_0^{t_1} P_{_{ij}}(t) \, dt$.  The LT of this integral can easily be expressed as:

\begin{equation}
\frac{1}{s^2}\Big(\textbf{I}- \tilde{\textbf{q}}(s) \Big)^{-1} \Big(\textbf{I}- \tilde{\textbf{h}}(s) \Big).
\label{eq1}
\end{equation}
Therefore when Equation (\ref{eq1}) is inverted at time $t_1$ we obtain $\int_0^{t_1} P_{_{ij}}(t) \, dt$, which is the expected amount of time spent in state $j$ from time $t=0$ to $t=t_1$ (assuming the process began in state $i$).


\subsection{First passage distributions}

Other important items available from an SMP are first passage distributions, $G_{_{ij}}(t)$. In the reliability example this would give us, for example, the distribution of the time to reach the ``Unrepairable" state. $\textbf{G}(t)$ is the matrix valued function of the CDFs of the first passage times from state $i$ to state $j$, and $\textbf{g}(t)$ is the corresponding matrix of PDFs.  $G_{_{ij}}(\infty)$ could be less than one if there is positive probability that the process may not ever reach state $j$ from state $i$.  These distributions are useful in determining how likely transitions are, and how long they may take.  They are also useful for calculating other quantities of interest.  From Equation (4.6) in \cite{pyke2} we obtain

\begin{equation}
\tilde{\textbf{g}}(s) = \tilde{\textbf{q}}(s) \Big(\textbf{I}- \tilde{\textbf{q}}(s) \Big)^{-1} \bigg[\textbf{I} \circ \Big(\textbf{I}- \tilde{\textbf{q}}(s) \Big)^{-1} \bigg]^{-1}
\label{eqnfirstpas}
\end{equation}
and we know,
\begin{equation}
\tilde{\textbf{G}}(s) = \frac{1}{s} \tilde{\textbf{g}}(s)  .
\label{eqnfirstpas2}
\end{equation}

At first glance Equation (\ref{eqnfirstpas}) may look intimidating, but for any matrix $\textbf{A}$, the operation $\textbf{I} \circ \textbf{A}$ just zeros out all the non-diagonal elements of $\textbf{A}$.  Then the inverse of a diagonal matrix is just the matrix of reciprocals of the diagonal elements (if they are all nonzero).  So once $\big(\textbf{I}- \tilde{\textbf{q}}(s) \big)^{-1}$ is calculated, the rest of Equation (\ref{eqnfirstpas}) is simple matrix manipulation.

For processes with an absorbing state, such as Figure \ref{figRelExamp}, we may be interested in finding the first passage hazard function.  In its basic form the hazard function is defined as:

\begin{equation}
\lambda(t) = \frac{f(t)}{1-F(t)} \, .
\label{eqn2}
\end{equation}

For the SMP in Figure \ref{figRelExamp}, the hazard function of the first passage can be calculated by finding $G_{_{1,3}}(t)$, $g_{_{1,3}}(t)$, and using Equation (\ref{eqn2}).

The first passage distributions can be used to find other quantities, as we see in the next section.

\subsection{Markov renewal process probabilities}

The probability $v_{_{ij}}(k;t)$ that the process has reached a particular state a certain number of times, is used when the Markov renewal process viewpoint is taken.  It provides information about the probability that state $j$ has been visited exactly $k$ times by the process up to time $t$.  Again considering the reliability example in Figure \ref{figRelExamp}, this would tell us how probable it is that 10 repairs would be needed by a particular time.  The cumulative version, how probable it is that 10 or fewer repairs would be needed by a particular time, is given by $V_{_{ij}}(k;t)$. The formulas are simpler if not presented in matrix form.  The first of these is Equation (5.4) of \cite{pyke2}:

\begin{equation}
  \tilde{v}_{_{ij}}(k;s) = \left\{
  \begin{array}{l l}
    \frac{1}{s}\Big( \tilde{g}_{_{ij}}(s)  \Big[ 1- \tilde{g}_{_{jj}}(s) \Big]  \Big[ \tilde{g}_{_{jj}}(s) \Big]^{(k-1)}    \Big) & \quad \text{if $k  \in \{1,2,3,\dots\}$}\\
    \frac{1}{s}\Big( 1- \tilde{g}_{_{ij}}(s)   \Big) & \quad \text{if $k=0$}\\
  \end{array} \right.
\end{equation}

Since these are probabilities of mutually exclusive events, we can add them to find the cumulative probability.  We find that $\sum_{n=0}^{k} v_{_{ij}}(n;s)$ is a telescoping sum and obtain:

\begin{equation}
  \tilde{V}_{_{ij}}(k;s) = \frac{1}{s}\Big( 1- \tilde{g}_{_{ij}}(s) \Big[ \tilde{g}_{_{jj}}(s) \Big]^{k} \Big) ,\text{ for } k  \in  \{0,1,2,\dots\}   \, .
\label{eqn13}
\end{equation}
In matrix form this equation is:
\begin{equation}
  \tilde{\textbf{V}}(k;s) = \frac{1}{s}\Big( \textbf{1}-\tilde{\textbf{g}}(s) \circ \Big[ \textbf{1} \big( \big[ \textbf{I} \circ  \tilde{\textbf{g}}(s) \big]^{k} \big) \Big] \Big) ,\text{ for } k  \in  \{0,1,2,\dots\}  \, .
\end{equation}

Presenting these probabilities in terms of first passage times has an intuitive interpretation:  $V_{_{ij}}(k;t)$ is just $1$ minus the CDF of the first passage from state $i$ to state $j$ convolved with the first passage CDF from state $j$ to $j$, $k$ times.  In other words the probability that we must repair the system $k$ times or less is $1$ minus the CDF of reaching the repair state for the first time convolved with reaching it $k$ more times.

The quantities discussed in this section can be very useful for the planning and allocation of resources.  Viewing $v_{_{ij}}(k;t)$ as the probability mass function of the random variable $K$, the next section discusses the expectation of $K$.

\subsection{Expected values of the Markov renewal process}

$M_{_{ij}}(t)$, the expected number of visits to state $j$ at time $t$, is just the expectation of $K$ or $\sum_{k=0}^{\infty} k \, v_{_{ij}}(k;t)$.  This also can be presented in terms of first passage distributions, but is more elegant as presented in Equation (5.11) from \cite{pyke2}:

\begin{equation}
\tilde{\textbf{M}}(s) = \frac{1}{s} \Big[  \Big(\textbf{I}- \tilde{\textbf{q}}(s) \Big)^{-1}  -    \textbf{I}    \Big] \, .
\label{eqn3}
\end{equation}

$\textbf{M}(t)$ is what defines a Markov renewal process (MRP) and Equation (\ref{eqn3}) is a theoretical link between SMPs and MRPs.  Roughly it says that given a SMP, using Equation (\ref{eqn3}) we can find a corresponding MRP and vice-versa.

\subsection{Asymptotic state probabilities}

The final quantities we discuss are the asymptotic state probabilities $\pi_{_{ij}}$.  In some instances $\pi_{_{ij}}$ may not be well defined; this can occur if the parameterization of a transition distribution has an infinite first moment (e.g., the L\'{e}vy distribution).

According to Equation (7.5) of \cite{pyke2}, we have for continuous time distributions:

\begin{equation}
\pi_{_{ij}} = \lim_{t\to\infty} P_{_{ij}} (t) = \frac{G_{_{ij}}(\infty) \int_{0}^{\infty} x h_{_{jj}}(x) \, dx}{\int_{0}^{\infty} x g_{_{jj}}(x) \, dx} \, .
\label{eqn4}
\end{equation}

In words, Equation (\ref{eqn4}) says that the asymptotic state probability is equal to the probability of reaching state $j$ from $i$, multiplied by the average time in state $j$, and divided by the average time to return to state $j$, if just arriving in state $j$.  If the expectations of the transition distributions are known, then $\int_{0}^{\infty} x h_{_{jj}}(x) \, dx$ is simple to calculate.  Usually $G_{_{ij}}(\infty)$ can be approximated by finding $G_{_{ij}}(\alpha)$, for $\alpha$ a relatively large number.  The main difficultly is finding  the mean of the first passage distribution, $\int_{0}^{\infty} x g_{_{jj}}(x) \, dx$.  If $G_{_{jj}}(\infty)=1$, one way to find the expected transition time of $g_{_{jj}}$ is to use the moment property of the LT, by calculating $-\frac{d}{ds}\tilde{g}_{_{jj}}(s)\big|_{s=0}$.  If $G_{_{jj}}(\infty)<1$, then $\int_{0}^{\infty} x g_{_{jj}}(x) \, dx = \infty$, and Equation (\ref{eqn4}) no longer holds; however, if $j$ is an absorbing state $\pi_{_{ij}} = G_{_{ij}}(\infty)$, and if $j$ is a transient state, $\pi_{_{ij}} = 0$.



Having introduced the primary equations for SMPs, we now proceed with an application.  The difficulties in applying the equations from \cite{pyke2} are the computations involved in passing to and from the LT domain, and the calculation of $\big(\textbf{I}- \tilde{\textbf{q}}(s) \big)^{-1}$.

\section{Application: Movement of Coronary Patients}\label{sectApplication}

This section focuses on applying the theory from the previous sections.  To illustrate, we use a semi-Markov process from \cite{Kao:1} that models the movement of coronary patients in a hospital, specifically the myocardial infarction (heart attack) positive patients.  This general type of model for patient movement is of interest to hospital administrators for planning and allocation of physical and personnel resources.  However, it is quite analogous to other logistic problems such as the flow of aircraft through the depot maintenance process as discussed in \cite{Huz:6}.

Inferential analysis of the movement of coronary patients, such as the selection of transition distributions and parameter estimation from the data, can be found in \cite{Kao:1}.  For this data maximum likelihood estimation was used, but Bayesian inference and other methods could also be implemented. We do not discuss the inference in this example in order to focus on solving the SMP after parameter estimates are obtained.

Coronary patients are assumed to be in one of nine possible states: coronary care unit (CCU), post-coronary care unit (PCCU), intensive-care unit (ICU), medical unit (MED), surgery (SURG), ambulatory care (AMB), extended care facility (ECF), HOME, and DIED.  The last three states are considered to be absorbing.

For the matrices below, the rows represent the instance of starting the process in a particular state and the columns present a quantity of a state given that row. The probability transition matrix of the process, $\textbf{p}$, is:

\[
\textbf{p}=
\bordermatrix{~&\text{CCU}&\text{PCCU}&\text{ICU}&\text{MED}&\text{SURG}&\text{AMB}
               &\text{ECF}&\text{HOME}&\text{DIED} \cr
\text{CCU}& 0.0000 & 0.7447 & 0.0084 & 0.1339 & 0.0042 & 0.0063 & 0.0000 & 0.0063 & 0.0962 \cr
\text{PCCU}& 0.0192 & 0.0000 & 0.0137 & 0.0247 & 0.0027 & 0.0027 & 0.0577 & 0.8298 & 0.0495 \cr
\text{ICU}& 0.0000 & 0.5833 & 0.0000 & 0.1667 & 0.0833 & 0.0000 & 0.0000 & 0.0000 & 0.1667 \cr
\text{MED}& 0.0000 & 0.0135 & 0.0405 & 0.0000 & 0.0135 & 0.0270 & 0.0811 & 0.7028 & 0.1216 \cr
\text{SURG}& 0.0000 & 0.0000 & 0.0000 & 0.0000 & 0.0000 & 0.0000 & 0.0000 & 1.0000 & 0.0000 \cr
\text{AMB}& 0.0000 & 0.0000 & 0.0000 & 0.0000 & 0.0000 & 0.0000 & 0.0000 & 1.0000 & 0.0000 \cr
\text{ECF}& 0.0000 & 0.0000 & 0.0000 & 0.0000 & 0.0000 & 0.0000 & 0.0000 & 0.0000 & 0.0000 \cr
\text{HOME}& 0.0000 & 0.0000 & 0.0000 & 0.0000 & 0.0000 & 0.0000 & 0.0000 & 0.0000 & 0.0000 \cr
\text{DIED}& 0.0000 & 0.0000 & 0.0000 & 0.0000 & 0.0000 & 0.0000 & 0.0000 & 0.0000 & 0.0000 \cr
}.
\]

Waiting time distributions for the process are assumed to be Weibull, as defined below. If $X$ has a Weibull($\gamma$,$\theta$) the PDF of $X$ is:

\begin{equation}
f_{_{X}}(x) = \frac{\gamma}{\theta} x^{\gamma-1}e^{-(x^{\gamma}/\theta)}
\end{equation}

The transition distributions as given in \cite{Kao:1} are:
\begin{itemize}
\item $f_1$ is Weibull$(4.738025, 4566277818.13)$
\item $f_2$ is Weibull$(2.207438, 14541.6089)$
\item $f_3$ is Weibull$(0.766338, 16.6991)$
\item $f_4$ is Weibull$(2.303331, 1017649.5158)$
\item $f_6$ is Weibull$(1.623492, 4707.3132)$
\end{itemize}

The transition matrix $\textbf{f}(t)$ is:
\[
\textbf{f}=
\bordermatrix{~&\text{CCU}&\text{PCCU}&\text{ICU}&\text{MED}&\text{SURG}&\text{AMB}
               &\text{ECF}&\text{HOME}&\text{DIED} \cr
\text{CCU} & 0 & f_1 & f_1 & f_1 & f_1 & f_2 & 0 & f_2 & f_3 \cr
\text{PCCU} & f_4 & 0 & f_1 & f_4 & f_1 & f_1 & f_4 & f_4 & f_6 \cr
\text{ICU} & 0 & f_4 & 0 & f_1 & f_1 & 0 & 0 & 0 & f_3 \cr
\text{MED} & 0 & f_4 & f_4 & 0 & f_4 & f_4 & f_4 & f_4 & f_6 \cr
\text{SURG} & 0 & 0 & 0 & 0 & 0 & 0 & 0 & f_4 & 0 \cr
\text{AMB} & 0 & 0 & 0 & 0 & 0 & 0 & 0 & f_4 & 0 \cr
\text{ECF} & 0 & 0 & 0 & 0 & 0 & 0 & 0 & 0 & 0 \cr
\text{HOME} & 0 & 0 & 0 & 0 & 0 & 0 & 0 & 0 & 0 \cr
\text{DIED} & 0 & 0 & 0 & 0 & 0 & 0 & 0 & 0 & 0 \cr
}.
\]

With $n=9$, $\textbf{p}$, and $\textbf{f}(x)$ we have defined an SMP.  For a given time $t$ we want to find some quantity, for example, the value of the first passage CDF from state $1$ to state $3$ at time $t=10$ i.e., $G_{_{1,3}}(10)$. First we compute $\tilde{f}(s)$ at the points needed by EULER, by elementwise numeric integration since the distributions do not have closed-form LTs.
Then using Equations (\ref{eqnfirstpas}) and (\ref{eqnfirstpas2}), we calculate $\tilde{\textbf{G}}(s)$.  Finally, the EULER algorithm gives an estimate of $\textbf{G}(10)$, from which we take the element on the first row and third column.  The computation time to calculate the matrix $\textbf{G}(10)$ is less than a second.  Similar methods are demonstrated in more detail in \cite{WarrHuz1}.

For processes with only a few states, we can afford to compute $(\textbf{I}-\textbf{q})^{-1}$ many times with only a small computational penalty; for processes with many states it would be more efficient to compute the inverse symbolically once, using a computer algebra package such as Mathematica or Maple, at the beginning of the calculations.

The time consuming calculation for this SMP is finding $\tilde{\textbf{f}}(s)$.  To reduce the computation time we only find this once for each time point $t$ and make all of the calculations from it before discarding the values.  Programming this way makes the code more complex, but speeds up the computations.

\begin{figure}[ht]
\begin{center}
\includegraphics[width=5in]{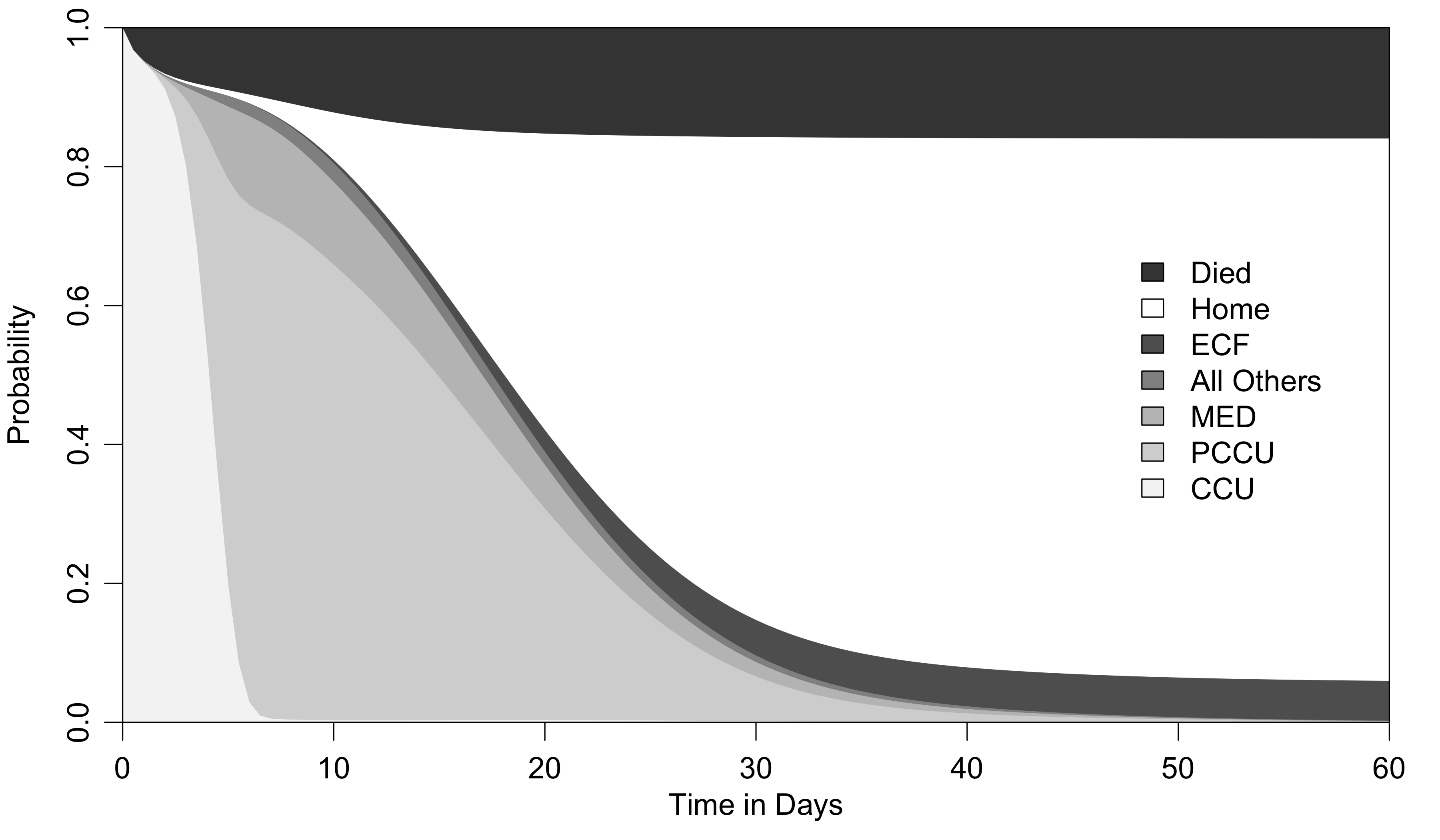}
\caption{The probabilities of being in a state at a particular time, given the patient began in the CCU state.}
\label{figTransStateProbs}
\end{center}
\end{figure}

For this SMP we calculated the transition probability matrix $\textbf{P}(t)$ at 120 points, spaced at 12 hours apart.  If we assume the patient began in state CCU we find the following probabilities in Figure \ref{figTransStateProbs}.  At any given time (a vertical line), the proportion of color of the line represents the probability a patient is in the corresponding state.  Clearly as time progresses the patient is absorbed into ECF, HOME or DIED.

We also obtain first passage distributions from these calculations;  from Equations (\ref{eqnfirstpas}) and (\ref{eqnfirstpas2}) we obtain both $\textbf{g}(t)$ and $\textbf{G}(t)$.  Using these we can find a conditional first passage hazard using the equation

\begin{equation}
\lambda_{_{ij}}(t) = \frac{g_{_{ij}}(t)}{G_{_{ij}}(\infty)-G_{_{ij}}(t)} \, .
\label{eqn10}
\end{equation}

\begin{figure}[ht]
\begin{center}
\includegraphics[width=6in]{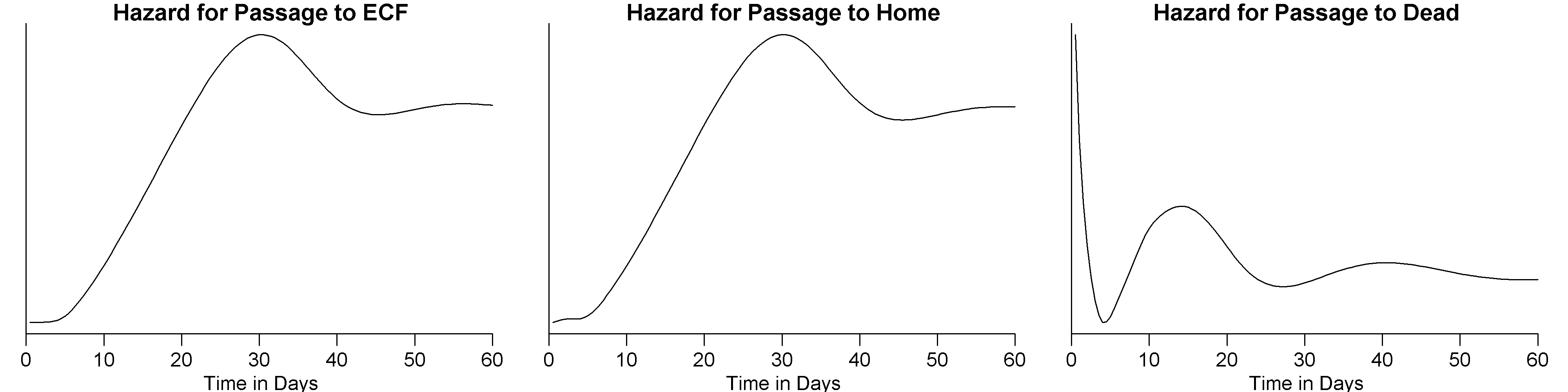}
\caption{The conditional hazard functions for the three absorbing states if the process began in state CCU.}
\label{figHazard}
\end{center}
\end{figure}

In Equation (\ref{eqn10}) the usual $1$ in the denominator has been replaced with $G_{_{ij}}(\infty)$.  This makes the assumption that the process eventually ends in state $j$.  Therefore Equation (\ref{eqn10}) gives a conditional hazard function, given the process reaches state $j$.  Figure \ref{figHazard} shows the conditional hazard functions for the three absorbing states (again assuming the process began in state CCU).

Next we look at some of the functions of the the Markov renewal process associated with this SMP.  The probability of revisiting a state in this SMP is small, so we only consider a few values of $k$ for $\textbf{v}(k;t)$.  We focus on $\textbf{v}(k;t)$ since $\textbf{V}(k;t)$ is easily calculated from $\textbf{v}(k;t)$.  For the interesting cases, the probabilities of not visiting a state before $t=60$, $\textbf{v}(0;60) = P(N_{_{j}}(60)=0|Z(0)=i)$, are given in the following matrix, where ``60" is 60 days.

\[
\textbf{v}(0;60) =
\bordermatrix{~&\text{CCU}&\text{PCCU}&\text{ICU}&\text{MED}&\text{SURG}&\text{AMB}
               &\text{ECF}&\text{HOME}&\text{DIED} \cr
\text{CCU} & 0.9854 & 0.2454 & 0.9749 & 0.8420 & 0.9894 & 0.9872 & 0.9426 & 0.2191 & 0.8405 \cr
\text{PCCU}& 0.9806 & 0.9766 & 0.9848 & 0.9698 & 0.9955 & 0.9963 & 0.9385 & 0.1214 & 0.9412 \cr
\text{ICU} & 0.9886 & 0.4105 & 0.9844 & 0.8158 & 0.9112 & 0.9933 & 0.9506 & 0.2795 & 0.7772 \cr
\text{MED} & 0.9993 & 0.9627 & 0.9593 & 0.9922 & 0.9829 & 0.9727 & 0.9163 & 0.2186 & 0.8687 \cr
\text{SURG}& 1.0000 & 1.0000 & 1.0000 & 1.0000 & 1.0000 & 1.0000 & 1.0000 & 0.0000 & 1.0000 \cr
\text{AMB} & 1.0000 & 1.0000 & 1.0000 & 1.0000 & 1.0000 & 1.0000 & 1.0000 & 0.0000 & 1.0000 \cr
}
\]

The higher the probability in the above matrix, the less likely that the state will be visited.  The next matrix shows $P(N_{_{j}}(60)=1|Z(0)=i)$ or $\textbf{v}(1;60)$ for the non-absorbing states.

\[
\textbf{v}(1;60) =
\bordermatrix{~&\text{CCU}&\text{PCCU}&\text{ICU}&\text{MED}&\text{SURG}&\text{AMB}
               &\text{ECF}&\text{HOME}&\text{DIED} \cr
\text{CCU} & 0.0144 & 0.7370 & 0.0248 & 0.1568 & 0.0106 & 0.0128 & 0.0574 & 0.7809 & 0.1595 \cr
\text{PCCU}& 0.0191 & 0.0229 & 0.0150 & 0.0300 & 0.0045 & 0.0037 & 0.0615 & 0.8786 & 0.0588 \cr
\text{ICU} & 0.0112 & 0.5760 & 0.0153 & 0.1828 & 0.0888 & 0.0067 & 0.0494 & 0.7205 & 0.2228 \cr
\text{MED} & 0.0007 & 0.0365 & 0.0401 & 0.0077 & 0.0171 & 0.0273 & 0.0837 & 0.7814 & 0.1313 \cr
\text{SURG}& 0.0000 & 0.0000 & 0.0000 & 0.0000 & 0.0000 & 0.0000 & 0.0000 & 1.0000 & 0.0000 \cr
\text{AMB} & 0.0000 & 0.0000 & 0.0000 & 0.0000 & 0.0000 & 0.0000 & 0.0000 & 1.0000 & 0.0000 \cr
}
\]

This matrix shows which states are likely to be visited once, the higher the probability the more likely to be visited.  This matrix clarifies that even though the process began in a state $i$, $N_{_{i}}(t)=0$ until it is visited again.  Therefore, $N_{_{i}}(t)$ is counting the number of times the process transitioned to state $i$ after $t=0$, not the number of times the process has been in state $i$.

The next matrix shows $\textbf{v}(2;60)$ for all the relevant cases.  Clearly, it is quite unlikely that any of the states will have 2 visits from the same patient.

\[
\textbf{v}(2;60) =
\bordermatrix{~&\text{CCU}&\text{PCCU}&\text{ICU}&\text{MED} \cr
\text{CCU} & 0.00020 & 0.01722 & 0.00038 & 0.00120 \cr
\text{PCCU}& 0.00028 & 0.00051 & 0.00023 & 0.00022 \cr
\text{ICU} & 0.00013 & 0.01327 & 0.00022 & 0.00138 \cr
\text{MED} & 0.00001 & 0.00074 & 0.00061 & 0.00005 \cr
}
\]

Next we show the expected number of visits, $M_{ij}(t)$, at 60 days, or $\textbf{M}(60)$.  We display only the values if starting in a non-absorbing state.

\[
\textbf{M}(60)=
\bordermatrix{~&\text{CCU}&\text{PCCU}&\text{ICU}&\text{MED}&\text{SURG}&\text{AMB}
               &\text{ECF}&\text{HOME}&\text{DIED} \cr
\text{CCU} & 0.015 & 0.773 & 0.026 & 0.159 & 0.011 & 0.013 & 0.057 & 0.781 & 0.159 \cr
\text{PCCU}& 0.020 & 0.024 & 0.015 & 0.030 & 0.005 & 0.004 & 0.061 & 0.879 & 0.059 \cr
\text{ICU} & 0.011 & 0.603 & 0.016 & 0.186 & 0.089 & 0.007 & 0.049 & 0.720 & 0.223 \cr
\text{MED} & 0.001 & 0.038 & 0.041 & 0.008 & 0.017 & 0.027 & 0.084 & 0.781 & 0.131 \cr
\text{SURG}& 0.000 & 0.000 & 0.000 & 0.000 & 0.000 & 0.000 & 0.000 & 1.000 & 0.000 \cr
\text{AMB} & 0.000 & 0.000 & 0.000 & 0.000 & 0.000 & 0.000 & 0.000 & 1.000 & 0.000 \cr
}
\]

This shows that all the states have a fairly low expected visit rate, and most patients would transition to an absorbing state with a relatively low number of transitions.

The final quantities we obtain are the asymptotic state probabilities.  These are fairly trivial when the process has one or more absorbing states.  When this is the case we only need to find $\textbf{G}(\infty)$.  The asymptotic probabilities of beginning in states $1$ through $6$ and being absorbed in states $7$ through $9$ are:

\[
\bm{\pi}=
\bordermatrix{~&\text{ECF}&\text{HOME}&\text{DIED} \cr
\text{CCU}&  0.0575 & 0.7830 & 0.1595 \cr
\text{PCCU}& 0.0615 & 0.8796 & 0.0589 \cr
\text{ICU}&  0.0499 & 0.7272 & 0.2229 \cr
\text{MED}&  0.0840 & 0.7846 & 0.1314 \cr
\text{SURG}& 0.0000 & 1.0000 & 0.0000 \cr
\text{AMB}&  0.0000 & 1.0000 & 0.0000 \cr
}.
\]

Each row adds to $1$, because the probability of reaching an absorbing state at infinity should be $1$ if the SMP has properly defined distributions for each transition.

In this example we demonstrated the capabilities of using modern computing power and techniques with the elegant theory presented in \cite{pyke2}.  All the computations were done in R and even with transitions that did not have closed-form LTs, the total computation time was about 70 seconds (on a computer with a dual core 1.9 GHz processor).  The computational time would dramatically decrease if the SMP was parameterized with transition distributions that had closed-form LTs.

We have calculated several quantities that are of use to statistical modelers and are not often included in an analysis.  Many times only the asymptotic probabilities are reported but these are of little use for processes that are not recurrent.

\section{Discussion and conclusions}

In this paper we have presented computational techniques for deriving quantities of interest in semi-Markov processes with smooth transition distributions; these techniques can be used in most continuous time SMPs. This section describes some performance issues and significant extensions which broaden the scope and applicability of SMPs.

\subsection{Performance}
SMPs with many states appear in applications such as queuing problems, where the number of states could reach more than 100,000.  The methodology we presented works well for SMPs with a small number of states, however at some point the number of states will become computationally problematic.  The computational performance will be affected by a several factors such as the sparseness of the $\textbf{q}(t)$ matrix, the number of states, $n$, and the number of elements in $\tilde{\textbf{q}}(s)$ that must be found numerically.  In SMPs with a small number of states, we can afford to find $(\textbf{I}- \tilde{\textbf{q}}(s))^{-1}$ many times, but as $n$ increases it may be beneficial to find it once symbolically using programs such as Maple or Mathematica.  To calculate SMPs with large $n$, care must be taken in defining states, parameterizing the state transitions and choosing methods for calculation of $(\textbf{I}- \tilde{\textbf{q}}(s))^{-1}$.

\subsection{Uncertainty quantification}

Besides solving for quantities such as $P_{_{ij}}(t)$ and $G_{_{ij}}(t)$, we are often interested in estimating the uncertainty associated with these quantities, based on uncertainties in estimating transition probabilities and waiting time distributions. A full treatment is beyond the scope of this paper, but we provide a summary here; see \cite[Chapt. 5]{flowgraphs} for further details and examples.

Writing the waiting time distributions as $F_{ij}(t; \theta_{ij})$, if the parameters $\theta_{ij}$ are estimated from sampled waiting times, a frequentist uncertainty analysis that accounts for sampling variability can be done as follows: draw bootstrap resamples for each transition, re-estimate the parameters for each $F_{ij}(t; \theta_{ij})$, and solve the process. After repeating this, say, $m$ times, the results can be used to construct confidence bands for quantities of interest.

Alternatively, we can do a Bayesian analysis by assigning prior distributions to the $\theta$'s, and computing posterior distributions based on waiting time samples. Then solving the process repeatedly using $m$ samples from the posterior of the $\theta$'s we obtain  $m$ estimates of, say, $G_{_{ij}}(t)$.  Using these $m$ estimates we can (point-wise in time) find 95\% credible intervals for the true $G_{_{ij}}(t)$.

\subsection{Nonparametric and semiparametric analysis}
The methods we have described may be applied to solve SMPs nonparametrically,
using sampled waiting-time data without assuming parametric distributions.
Given a sample $x_1, x_2, \ldots , x_n$ of waiting times for a given transition,
we use the empirical distribution function (EDF) $F_n(t)$ and the
Laplace-Stieltjes transform to define the empirical Laplace transform (ELT)
of the data as
\[
\hat{f}(s) = \int_0^\infty e^{-sx}dF_n(x) =
  \frac{1}{n}\sum_{i=1}^n e^{-s x_i}.
\]
After substitution of $\hat{f}_{ij}(s)$ for $\tilde{f}_{ij}(s)$ in the
construction of the $\tilde{\textbf{q}}$ matrix, all the algorithms described
in Section \ref{sectTheory} can be used without change to derive, e.g., the
empirical transform of a first passage distribution; it can be shown
\cite[]{CollinsDis} that the resulting ELTs are strongly consistent estimators
of the corresponding exact transforms. Parametric LTs and ELTs may also be mixed in
constructing $\tilde{\textbf{q}}$, for a semiparametric analysis.

Since the EDFs underlying ELTs are not smooth, the EULER method described in
Section \ref{sectTransformInversion} performs poorly. However, a modified EULER is developed in \cite{Collins:3}, which is both fast and accurate
(for large enough samples) for ELT inversion.

\vspace{.2in}\noindent
In conclusion, this paper fills a significant gap in the SMP literature by connecting the theory with functional formulas 
that can be used by practitioners in a large variety of applications.
Our ongoing work is directed towards expanding the applicability of the methods, improving performance, broadening the base of computational platforms, and developing software that is easier to use.

\section{Acknowledgments}
The authors thank Aparna Huzurbazar and Mike Hamada for their insightful comments.



\bibliographystyle{dcu}

\end{document}